\documentclass[lettersize,journal,onecolumn, 12pt,draftcls]{IEEEtran}

\usepackage{amsmath,amsfonts}
\usepackage{algorithmic}
\usepackage{algorithm}
\usepackage{array}
\usepackage[caption=false,font=normalsize,labelfont=sf,textfont=sf]{subfig}
\usepackage{textcomp}
\usepackage{stfloats}
\usepackage{url}
\usepackage{verbatim}
\usepackage{graphicx}
\usepackage{cite}
\usepackage{graphicx}
\usepackage{textcomp}
\usepackage{xcolor}
\usepackage{color}
\newcommand{\SJ}{\color{black}}
\newcommand{\MH}{\color{black}}

\newcommand{\AS}{\color{black}}
\newcommand{\AK}{\color{black}}

\usepackage{tikz} 
\usetikzlibrary{positioning}
\usepackage{multirow}
\newcommand{\tabitem}{{\textbullet}~}
\usepackage{array}
\usepackage{ragged2e}
\newcolumntype{M}[1]{>{\RaggedRight\hspace{0pt}}m{#1}}
\usepackage{booktabs}
\usepackage{float}
\usepackage{longtable}
\usepackage{lscape}
\usepackage{setspace}
\usepackage{makecell}

\usepackage{afterpage}
\usepackage{capt-of}
\begin{document}

\title{Open RAN xApps Design and Evaluation: Lessons Learnt and Identified Challenges}

\author{Marcin Hoffmann,~\IEEEmembership{Graduate Student Member,~IEEE,}
        Salim Janji, ~\IEEEmembership{Graduate Student Member,~IEEE,}
        Adam Samorzewski,~\IEEEmembership{Graduate Student Member,~IEEE,}
        Łukasz Kułacz,~\IEEEmembership{Member,~IEEE,}
        Cezary Adamczyk,
        Marcin Dryjański,~\IEEEmembership{Senior Member,~IEEE,}
        Pawel Kryszkiewicz,~\IEEEmembership{Senior Member,~IEEE,}        
        Adrian Kliks,~\IEEEmembership{Senior Member,~IEEE,}
        Hanna Bogucka,~\IEEEmembership{Senior Member,~IEEE,}
\thanks{
Authors are with Rimedo Labs and Institute of Radiocommunications, Poznan University of Technology, Poznan, POLAND. Corresponding author: M. Hoffmann, e-mail: marcin.hoffmann@put.poznan.pl or marcin.hoffmann@rimedolabs.com}}

\markboth{ IEEE Journal on Selected Areas in Communications }%
{Hoffmann \MakeLowercase{\textit{et al.}}: Open RAN xApps Design and Evaluation: Lessons Learnt and Identified Challenges}

\IEEEpubid{0000--0000/00\$00.00~\copyright~2023 IEEE}

\maketitle

\begin{abstract}
The concept of open radio access networks (RAN) creates numerous opportunities for the development of various fields of economy. At the same time, a flexible and modular approach in the disaggregated RAN entails the need for careful design of both the overall RAN architecture and the process of implementation and deployment of new applications. It is assumed that the latter may be delivered by dedicated and specialized software companies. To make the whole process efficient, safe, and reliable, a joint effort among different sectors (industry, academia, standardization bodies) has to be guaranteed. Here, one of the important driving forces origins from the open-source community that often stimulates the development of a specific technology. In this paper, we address the challenges that have to be faced by third-party application developers in the context of Open RAN. Based on many implemented applications (called xApps or rApps), we compared various available solutions and posed the most critical issues that have to be tackled in the near future to stimulate the progress in open RAN development further. In particular, we start by comparing available open platforms for xApp development and testing. We present the details of implementing four selected applications describing the problems encountered. {\AK It is split into two logical parts - first, we identify the key ambiguities related to the development of new xApps, which address more complicated use cases like beam management. In the second part, we present the challenges related to the detailed implementation of software in existing open platforms. In the first case, we showed that dedicated beam mobility management xApp can lead to the reduction of beam switches and can keep beam failures at a low level. However, it requires access to detailed localization information. Similarly, the signaling storm detection xApp provides expected performance under the assumption that there is access to detailed information on, e.g., time advance resolution parameter. We concluded here that several aspects are still not well-defined to allow smooth software implementation; these include the rules for data reporting in time, parameters available in service models, and localization features. Concerning the second logical part, related to low-level implementation, we compared the numerical results of the traffic steering and quality-of-service-based resource allocation xApps and drew conclusions related to implementation and testing. In particular, we pointed out problems associated with the simulator itself, with the software, and conflicts inside.} Finally, we identify the key challenges which should be treated as incentives for joint academia-industry cooperation in the field of Open RAN. Thus, the paper presents the lesson learned during the first years of xApp development. 
\end{abstract}

\begin{IEEEkeywords}
Open RAN, 5G, 6G, xApp, ML
\end{IEEEkeywords}

\section{Introduction}
\IEEEPARstart{D}{isaggregation}, openness, flexibility, and modality - these are the new paradigms attributed to the next generation of wireless communication networks. Contrary to the traditional and prevalent approach to the radio access network (RAN) design, where most of the RAN elements are provided by one vendor and are hidden in the \textit{black-box},  the concept of the Open RAN assumes that potentially multiple players provide dedicated RAN modules. Such a modular approach allows operators to modify and improve only selected network functionalities instead of completely replacing the black-boxed software. It is the network operator who decides what functions in the network should be activated or deactivated, which should be improved, kept unchanged, or uninstalled. All of these modifications can be done by means of proper manipulation of the installed software modules. This, in turn, opens the possibility for incremental system modifications following the concept of continuous integration and continuous development (CI/CD). The trend towards open and modular RAN is emphasized by the standardization activities related to Open RAN, which are led by the O-RAN ALLIANCE \cite{ORANALLIANCE}. The set of standards released by this organization specifies the overall Open RAN architecture, requirements, and functionalities. In particular, new and open interfaces have been proposed to incentivize xApp/rApp providers to implement and deliver new algorithms dealing with specific aspects of wireless communications. 
However, along with the numerous and evident benefits related to opening and disaggregating the RAN, there are significant challenges related to the practical implementation of such a vibrant concept. First, the way for implementation and deployment of new xApps/rApps has to be somehow unified and automated so that every interested software provider may deliver valuable contributions to the community. Next, opening the RAN part to numerous, often external providers causes various security issues which must be tackled carefully. Also, the coexistence of applications originating from different xApp/rApp providers may lead to potential consistency and confluence problems and prospective conflicts. These topics are now the subject of both academic and industrial debate. However, despite all the efforts put into the foundation of an open and disaggregated RAN environment, the technology is still in its early stage of development. The architecture, although precisely specified in O-RAN ALLIANCE standards, is still modified and being adjusted to address new challenges and to reply to the recent findings. Moreover, practical implementations also face many difficulties, including a lack of trusted simulation environments, commonly agreed ways for providing new software modules, ways of testing, and performance benchmarking. 

These problems are of particular importance from the perspective of the above-mentioned xApp/rApp providers, who still have limited possibilities of delivering new applications. When new xApp/rApp is being implemented, it has to be first simulated reliably and comprehensively, it has to be tested against numerous threats and risks, and the whole process has to be automated. Nowadays, it is not the case. In this paper, we address this niche by presenting the observations gained in the years of xApp/rApp design and implementation. By implementing some xApps/rApps of different kinds, types, and scopes of functionalities, we were able to discuss the current state of the development art from the perspective of the xApp/rApp provider. We present our lessons learned and gained experience to identify key challenges along with standardization and research directions. To avoid the promotion of any commercial solutions and to promote open science, we concentrate on the Open RAN applications prepared with the openly available software and mutually compare the achieved results.

{\AK The paper is structured as follows - it contains four logical sections. First, a concise review of what O-RAN is is provided; next, the existing implementation frameworks are discussed and compared; third, we present four original xApp implementation results, discussing their performance and drawing conclusions about the whole design process.} Its novelty can be summarized as follows:
\begin{itemize}
\item we present in detail four xApps, illustrating the message exchange between the particular O-RAN blocks,
\item we provide a detailed comparison of currently available software platforms and discuss their pros and cons,
\item we discuss the O-RAN architectural ambiguities based on the challenges that have been faced during the implementation of the xApps,
\item analogously, we share our observations in the context of current limitations related to xApp implementation.
\end{itemize}
To precisely reflect the above topics, this paper is split into seven chapters, where the following section recaps the O-RAN architecture, RAN Intelligent Controller, and proposed use cases. Chapter 3 overviews the open-source platform for xApp development and testing. Next, Chapters 4 and 5 present the ways of four xApp implementations and draw conclusions related to architecture ambiguities and practical implementation, respectively. Chapter 6 discusses the key research challenges. The whole work is summarized in Chapter 7. 

\section{O-RAN Architecture, RIC and Use Cases}
\textbf{O-RAN ALLIANCE} \cite{ORANALLIANCE} is the main standardization body specifying the O-RAN reference architecture, interfaces, deployment scenarios, use cases, etc. In addition to this, it also leads official plugfests, provides an open-source implementation of the O-RAN stack, and interoperability and testing of the O-RAN solutions. 

This chapter provides an overview of the O-RAN architecture as defined by O-RAN ALLIANCE with a special focus on the \textbf{RAN Intelligent Controller (RIC)} along with xApps.

\subsection{O-RAN Architecture}
\begin{figure}[htbp]
\centerline{\includegraphics[height=8.0cm]{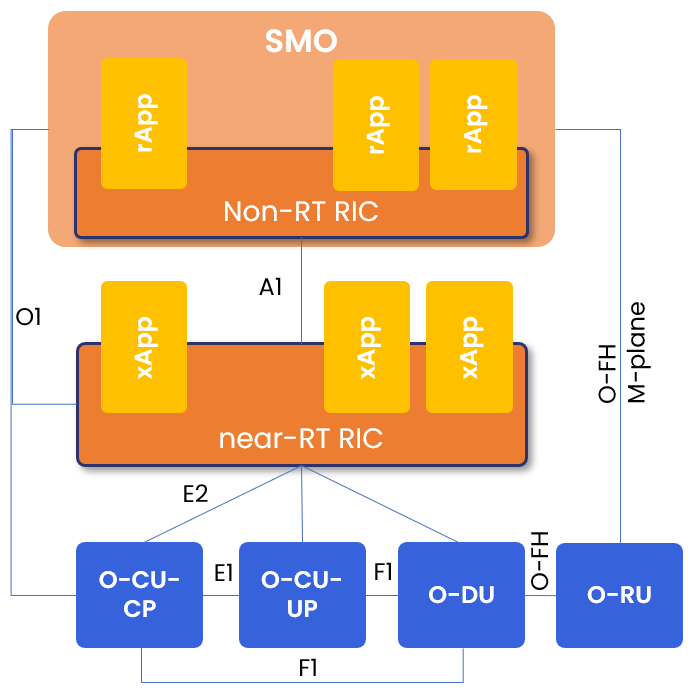}}
\caption{O-RAN architecture, as defined by O-RAN ALLIANCE}
\label{fig:oran-arch}
\end{figure}

The overall O-RAN architecture (see Figure~\ref{fig:oran-arch}) is defined within \cite{ORANarchitecture} and builds upon 3GPP RAN standards towards openness and intelligence by adopting RAN splits, new interfaces, RICs, and Service Management and Orchestration (SMO).

The O-RAN adopts split 2 (also referred to as higher-layer split, HLS), between PDCP and RLC protocols within the NR air interface stack; and split 7.2x (also referred to as lower-layer split, LLS), within the PHY layer. The corresponding elements of the RAN are called \textbf{O-RAN Central Unit (O-CU)}, \textbf{O-RAN Distributed Unit (O-DU)}, and \textbf{O-RAN Radio Unit (O-RU)}. 

O-CU is further split into the control plane (O-CU-CP), which covers Radio Resource Control (RRC) with Packet Data Convergence Protocol-Control Plane (PDCP-C) protocols, and the user plane (O-CU-UP) covering and Service Data Adaptation Protocol (SDAP) with PDCP-User Plane (PDCP-U). O-DU, in turn, encompasses Radio Link Control (RLC), Medium Access Control (MAC), and a high-physical layer, including the MAC scheduler. Finally, O-RU includes low-physical layer functionality like Orthogonal Frequency Division Multiple Access (OFDMA) processing, beamforming, and Radio Frequency (RF) front end.

An important element introduced in O-RAN is the RAN Intelligent Controller (RIC), a separated-out entity from the processing units that allow to access RRM functions. RIC is split onto \textbf{Non-Real-Time RIC (Non-RT RIC)} and \textbf{Near-Real-Time RIC (Near-RT RIC)}. The former works in the timescale of above 1 s, is used for non-real-time radio resource management, higher layer procedure optimization, and policy optimization in RAN, and enables the artificial intelligence (AI) and machine learning (ML) workflow for RAN components. In addition, it provides policy-based guidance for the applications in Near-RT RIC and delivers Enrichment Information (EI) for the Near-RT RIC's applications. Near-RT RIC, on the contrary, is part of the RAN to enable control and optimization of algorithms for radio resource management and it works with the control loop in a timescale of larger than 10 ms and less than 1 s utilizing the use-case specific applications called xApps.

O-RAN ALLIANCE also specifies new interfaces including Open Fronthaul (OFH), which connects O-DU to O-RU, E2, and A1 serving as control loop connections, and O1, O2, OFH M-plane - i.e. management interfaces. O-CU-CP, O-CU-UP, and O-DU are also referred to as "E2 Nodes" in the O-RAN architecture. This is because they are connected via the E2 interface to the Near-RT RIC, by which their functionality can be controlled through external applications, i.e., the abovementioned xApps.

Among the mentioned interface, E2 and A1 are considered important in this paper, namely:

\begin{itemize}
    \item \textbf{E2 interface}, which creates a closed loop within the RAN domain, is used to send the RIC control and policy toward E2 Nodes and to obtain the feedback from E2 Nodes to the Near-RT RIC.
    \item \textbf{A1 interface}, which is used to provide policies, EI, and ML models towards Near-RT RIC, and to get the policy feedback back to the Non-RT RIC.
\end{itemize}

\subsection{O-RAN Near-RT RIC, xApps and Use Cases}
Near-RT RIC serves as a software platform to allow the xApps to control the RAN. This is supported by the RAN and UE databases storing the network state, along with xApp management, security, and conflict mitigation functions. It enables near real-time control optimization of the E2 Nodes via actions sent over the E2 interface, including CONTROL, INSERT, POLICY, and REPORT services \cite{ORANarchitecture}. The detailed description of Near-Real-Time RIC is defined in \cite{nearRTRIC}. 

E2 Nodes mentioned above expose parameters and functionalities towards the RIC through the E2 interface, which can be used by xApps and rApps to tune the behavior of the radio network. Examples of xApps are mobility, interference or beamforming management, traffic steering, load balancing, slice control, admission control, signaling anomaly detection, etc.

In this paper, we focus on xApps which  are applications to run at the Near-RT RIC. An xApp provides information to the Near-RT RIC about the data types it consumes and about outputs it produces. Such application is, in principle, independent of the Near-RT RIC and may be provided by a third-party provider. An xApp controls a specific RAN functionality exposed by the E2 Node using the E2 service models (E2SM). The current service models include KPM (Key Performance Measurements), RC (RAN Control), NI (Network Interface), and CCC (Cell Configuration and Control) \cite{nearRTRIC}.

To summarize, Near-RT RIC is one of the key elements in the O-RAN architecture, which allows feeding intelligence into the operations of the RAN. It creates a platform on which the software providers could build per-use case RRM algorithms to allow the optimization of radio resources for specific scenarios, also known as use cases, which are covered in the next subsection. The use cases, based on which xApps and rApps are developed are defined in \cite{ORANUseCases} and are based on the requirements of O-RAN ALLIANCE members. Those requirements also come as input to the O-RAN ALLIANCE's standardization in the form of priorities from Telecom Infra Project, an organization, which brings together operators. TIP’s OpenRAN program supports the development of disaggregated and interoperable RAN solutions based on service provider requirements \cite{TIP}. Specifically, within the RRM part, TIP defines the RAN Intelligence and Automation subgroup (RIA) aiming to develop and deploy AI-based xApps for use cases like RRM, SON, or Massive MIMO, etc.

The current set of O-RAN ALLIANCE use cases, as specified in \, cite{ORANUseCases} covers 23 items and includes among others: V2X HO management, UAV radio resource allocation, QoE optimization, traffic steering, Massive MIMO BF optimization, RAN sharing, QoS-based resource optimization, RAN slice SLA assurance, Dynamic Spectrum Sharing, indoor positioning, signaling storm protection, congestion protection, energy saving, etc. 

Based on the use case definition and description, defined by O-RAN WG1, other working groups define parameters and procedures to create a normative way for interoperable interfaces allowing interworking between vendors. Examples include parameters and new service models at the E2 interface, or policy definitions for those use cases at the A1 interface.
 
\section{O-RAN Deployments - Open-source Platforms Comparisons}

To date, there are several open-source projects that are used in the implementation of Open RAN systems. Such platforms may provide the full stack that includes RAN software, RICs, SMO, or a subset of those components. This section presents several platforms along with a brief description of the modules they provide. Since a complete end-to-end deployment or simulation of 5G systems requires implementing both the RAN and Core Network (CN) domains we also mention the 5G CN implementation that each project leverages in its platform while discarding any LTE Evolved Packet Core (EPC) implementations. We also highlight a few differences between them and conclude with an evaluation of each platform based on the documentation they provide and the hands-on experience gained while testing some of these projects. 

\subsection{OpenAirInterface (OAI)\cite{oai_website}}\label{sec:oai_main}
OpenAirInterface (OAI) Software ALLIANCE (OSA) was established in 2014 by the non-profit organization EURECOM. Among others, OAI provides the following projects.

\subsubsection {OAI 5G CN and EPC CN}\label{sec:oai_cn}
these projects provide 5G standalone (SA) CN and 5G Non-standalone (NSA) CN network functions (NFs) implementations, respectively.

\subsubsection {OAI 5G RAN}\label{section:oai_5g_ran}
this OAI project implements software for both NSA and SA gNB, eNB, 5G NSA and SA UE, and LTE UE. 

\subsubsection {OAI's MOSAIC5G}\label{sec:mosaic5g}
this project develops control and orchestration frameworks on top of OAI's RAN and CN modules which allows for monitoring and controlling of the network. It includes Trirematrics and FlexCN platforms in its roadmap that provide SMO and CN control modules, respectively, and a FlexRIC software which we introduce below.

\textbf{E2 Agent and FlexRIC}: FlexRIC provides an SDK that can implement a multi-vendor O-RAN compliant RT RIC that is specialized for a certain service (e.g., slice control, traffic control, etc.) with built-in service models (SMs) and support for the creation of further SMs \cite{oai_flexric}. OAI's FlexRIC design is meant to be extensible and compact with minimum overhead. Furthermore, unlike RICs provided by other projects, it follows an event-driven approach rather than a poll-driven. The main modules contain an agent library that deploys E2-compatible agents in a base station and a server library that manages agents' connections, stores network information in the radio network information base (RNIB), and handles E2SM subscriptions. These subscriptions can be established by iApps which are controller internal applications that can either implement a specific control logic or expose E2SM subscriptions to xApps deployed on external controllers through different types of interfaces.

The agent library is radio access technology (RAT) and vendor-neutral which allows multi-RAT and multi-vendor deployments. Agents can also connect to multiple controllers through the server library which provides isolation between them. Furthermore, a virtualization layer with an agent can be implemented on top of a server deployment which allows recursive agent-server layers. This is beneficial in cases where we want to abstract out RAT heterogeneity or delegate control to multiple controllers per slice using different SMs. 





\subsection{O-RAN Software Community (OSC)\cite{osc_website}} \label{sec:osc_main}
OSC is founded by O-RAN and Linux Foundation and it aims to provide software that is fully O-RAN compliant. In general, the project encompasses all O-RAN-related components, RAN elements, and interfaces between them. 
We present some OSC projects below.


\subsubsection{O-DU} \label{sec:osc_du}
this project is composed of two sub-projects. \textit{O-DU} Low focuses on the baseband PHY reference design including three interfaces: L1/Fronthaul;\textit{O-DU Low}/\textit{O-DU High}, and \textit{O-DU Low}/accelerator. \textit{O-DU High}, is responsible for implementing L2 blocks for 5G NR SA mode that include NR MAC, NR Scheduler, and NR RLC layers. \textit{O-DU High} also provides \textit{DU APP} which configures and manages all O-DU operations, and interfaces with external entities (e.g., O-CU, RIC, etc.). Finally, it implements an O1 module to handle O1 communication. 

\subsubsection{O-CU}\label{sec:osc_cu}
\textit{O-CU} was supposed to provide O-CU UP. However, it seems the project was disbanded, and instead, OSC uses a binary test stub provided by Radisys for end-to-end testing. 

\subsubsection{Near-RT RIC}\label{sec:osc_nric}
this project provides an initial RIC platform to support xApps with limited support for O1, A1, and E2 interfaces.

\subsubsection{Non-RT RIC} \label{sec:osc_nonric}
in the context of Non-RT RIC, OSC provides a Non-RT RIC Control Panel which provides administrative and operator functions through A1 like policy management and Near-RT RICs setup. Also, an A1 Simulator module is implemented which terminates the A1 interface and allows testing the Non-RT RIC without the need for deployment of Near-RT RICs. To support management functions, an SMO project implements O1 and O1/VES interfaces that are responsible for the configuration, management, and report handling of NFs. Finally, and OAM project provides administrative and operator functions for O-RAN components.






\subsection{Open Networking Foundation (ONF)\cite{onf_website}}\label{sec:onf}
ONF was established as a project to develop software-defined networking (SDN) technologies and currently, it is driven by operators and a community of developers. ONF developed its SD-RAN project which provides a Near-RT RIC that was adapted to O-RAN specifications in its latest version at the time of writing this paper. Besides the Near-RT RIC which is called \textit{$\mu\text{ONOS-RIC}$} due to its implementation being based on ONF's \textit{\text{ONOS}} platform, SD-RAN provides open-source components for the control and user planes of CU and DU, a RAN simulator, and xApps development SDK. The CU/DU modules are derived from OAI's 5G RAN project (see \ref{section:oai_5g_ran}). SD-RAN leverages a microservice approach that is compatible with O-RAN specifications

\subsection{Open AI Cellular (OAIC) \cite{oaic_website}}\label{sec:srsran_oaic}
Founded by USA National Science Foundation, OAIC uses OSC’s Near-RT RIC (see \ref{sec:osc_nric}) on top of srsRAN \cite{srsran_website} which provides components for implementing a complete end-to-end 4G and 5G NSA networks. {\SJ For E2 implementation, OAIC leverages POWDER's E2 agents \cite{powder} in their architecture. Moreover, OAIC provides OAIC-T, an open-source AI cellular testing framework for testing xApps. It consists of a server that establishes the simulation environment according to input from configuration files, and the actors that perform the test actions received from the server. Each actor contains an AI core component and it can communicate with xApps or rApps under test, and srsUEs to generate radio testing signals.} Within its framework, srsRAN provides srsUE to deploy 4G/5G UEs {\SJ using ZeroMQ}, srsENB as an eNB implementation with 5G NSA support, and srsEPC as a lightweight implementation of LTE EPC, while it lacks an implementation of 5G CN (they advertise using Open5GS \cite{open5gs} for 5G CN). 

\subsection{OpenRAN Gym \cite{ORANGYM,Melodia2022,Melodia2022Transactions}}
{\SJ
Combining several software frameworks, OpenRAN Gym allows data acquisition of RAN performance indicators from emulators or testbeds and RAN control for testing of O-RAN compliant solutions powered by AI/ML. The platform encompasses the following.
\begin{itemize}
    \item Open experimental wireless platforms for acquiring RAN data and testing solutions (e.g., Colosseum which is the world's largest wireless network emulator, Arena testbed, etc.),
    \item RAN software implementations using srsRAN or OAI stacks,
    \item SCOPE framework which is used for data collection and control of RAN during run-time which also adds further networking and control functionalities (e.g., slicing) to the RAN software, and
    \item ColO-RAN provides a lightweight RIC adapted from OSC's RIC that allows for xApps/rApps to monitor KPMs and control the RAN.
\end{itemize}
Using these tools, solutions can be validated on the Colosseum emulator for example and then ported to heterogeneous testbeds seamlessly as described in \cite{ORANGYM} 
}

\subsection{Comparison of Different Platforms and Their Compatibility}
\begin{figure}[!h]
\centering
\begin{tikzpicture}[node distance={15mm}, thick, main/.style = {draw, rectangle}] 
\centering
\node[main, text width=2.5cm,align=center,color=white] (1) [fill={rgb, 255:red,243;green,168;blue,117}]{OAI's 5G CN (\ref{sec:oai_cn})}; 
\node[main, text width=2.5cm,align=center,color=white] (2) [right=0.2cm of 1, fill={rgb, 255:red,237;green,125;blue,49}]{Free5GC\cite{free5gc}};
\node[main, text width=2.5cm,align=center,color=white] (3) [right=0.2cm of 2, fill={rgb, 255:red,237;green,125;blue,49}]{Open5GS \cite{open5gs}};
\node[main, text width=2.5cm,align=center,color=white] (4) [below=1cm of 1, fill={rgb, 255:red,243;green,168;blue,117}]{OAI's 5G RAN (\ref{section:oai_5g_ran})}; 
\node[main, text width=2.5cm,align=center,color=white] (5) [right= 0.2cm of 4, fill=red!50]{srsRAN (\ref{sec:srsran_oaic})};
\node[main, text width=2.5cm,align=center,color=white] (6) [right= 0.2cm of 5, fill={rgb, 255:red,54;green,98;blue,219}]{OSC's RAN (\ref{sec:osc_du})};

\node[main, text width=2.5cm,align=center,color=white] (7) [below=2cm of 4, fill={rgb, 255:red,243;green,168;blue,117}]{OAI's FlexRIC (\ref{sec:mosaic5g})}; \node[main, text width=2.5cm,align=center,color=white] (8) [right= 0.2cm of 7, fill={rgb, 255:red,255;green,192;blue,0}]{ONF's $\mu\text{ONOS-RIC}$ (\ref{sec:onf})};
\node[main, text width=2.5cm,align=center,color=white] (9) [right= 0.2cm of 8, fill={rgb, 255:red,54;green,98;blue,219}]{OSC's RIC (\ref{sec:osc_nric} and \ref{sec:osc_nonric})};

\draw (1) -- (4);
\draw (2) -- (4);
\draw (3) -- (5);
\draw (4) -- (7);
\draw (5) -- node[midway, above left, sloped, pos=1] {\cite{oaic_website, nexran}~~~~} (9);
\draw (6) -- (9);
\draw (4) -- (8);
\draw (4) -- node[midway, above left, sloped, pos=1] {\cite{exhibition_osc_oairan}~~~~~~~~~~~}  (9);
\draw [-] (7) to [out=270, in=270, looseness=0.4] node[midway, above] {using RMR (see \ref{sec:osc_nric})} (9);
\end{tikzpicture} 
\caption{Different projects for building and testing a complete end-to-end 5G system with Open RAN functionalities and their compatibility. Starting from top, the first row lists 5G CN projects, the second row mentions 5G RAN implementations, and the last row lists RICs implementations. Colors indicate the vendor: OAI (light orange); OSC (blue); ONF (yellow); srsRAN (red) and other vendors (dark orange).}
\label{fig:projects_compatibility}
\end{figure}
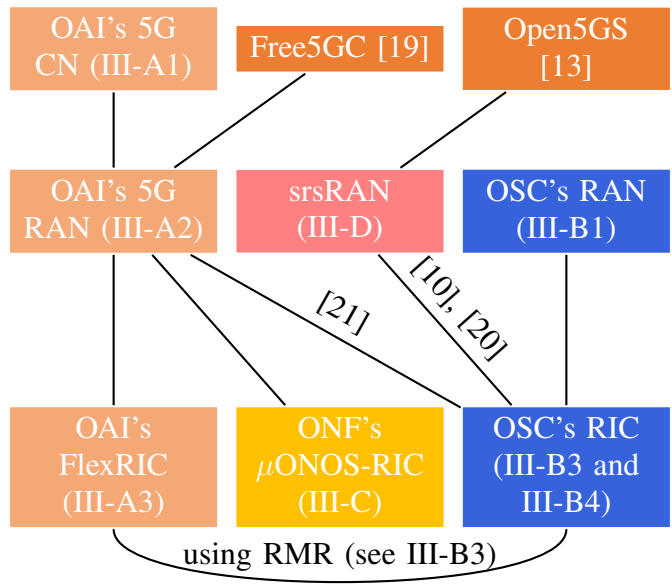

Table~\ref{tab:comparison} lists the perceived differences between the different implementation options. Furthermore, in Fig.~\ref{fig:projects_compatibility} we present the components used in currently available solutions and their combinations, and we also include other open-source CN projects that were not mentioned in our earlier discussion, which are compatible with some RAN implementations.

\begin{table}[hbtp]
\centering

    \caption{Comparison of open-source implementations along with exemplar xApps provided by each platform.}
    \label{tab:comparison}
    \begin{tabular}{p{.05\linewidth}M{.22\linewidth}M{.22\linewidth}M{.22\linewidth}M{.22\linewidth}}
        \toprule
& \textbf{OAI}& \textbf{OSC}& \textbf{ONF}& \textbf{ OAIC}\\\midrule
CN
&\tabitem OAI CN &\tabitem No CN or EPC &\tabitem OMEC CN &\tabitem LTE EPC \\\hline

RAN
&\makecell{\parbox{0.22\textwidth}{\tabitem Better CPU and memory utilization than srsRAN \cite{gringoli_performance_2018}\\ \tabitem Multiple UE simulation\\ \tabitem MOSAIC5G E2 agents \\\tabitem No O1 interface implementation}} 
&\makecell{\parbox{0.22\textwidth}{\tabitem Radisys CU lacks integration with open-source CU and RU implementations \\\tabitem DU and DU App \\\tabitem O1 interface}}
&\makecell{\parbox{0.22\textwidth}{\tabitem Leverages OAI's RAN modules \\\tabitem ONF's own RAN simulator with more features and capability to simulate a large number of UEs}}
&\makecell{\parbox{0.22\textwidth}{\tabitem Easier to modify \cite{gringoli_performance_2018}\\\tabitem POWDER\cite{powder} E2 agents within srsRAN stack\\\tabitem Single UE simulation \\\tabitem No F1 interface for CU/DU split\\\tabitem No O1 interface}}\\\hline

RIC
&\makecell{\parbox{0.22\textwidth}{\tabitem Better CPU, memory utilization, and latency than OSC's RIC\cite{oai_flexric}\\ \tabitem Recursive agent library for the abstraction of underlying topology \\\tabitem iApps have less overhead than xApps}} 
&\makecell{\parbox{0.22\textwidth}{\tabitem Completely O-RAN compliant  \\\tabitem All O-RAN components including Non-RT RIC \\\tabitem Requires more resources due to containerization and microservice structure}}
&\makecell{\parbox{0.22\textwidth}{\tabitem $\mu\text{ONOS-RIC}$ using ONOS modules \\\tabitem Code used in previous SDN activities and is therefore reliable \\\tabitem Good documentation \\\tabitem Latest version is fully O-RAN compliant}}
&\makecell{\parbox{0.22\textwidth}{\tabitem Uses OSC's RIC}}\\\hline
xApps
&\makecell{\parbox{0.22\textwidth}{\tabitem Key performance metrics (KPMs) monitoring, slice monitoring and control, and traffic controller}} 
&\makecell{\parbox{0.22\textwidth}{\tabitem Anomaly detection, HelloWorld xApp, HW-go xApp, KPM monitoring, QoE predictor, RIC APP ML, RIC Measurement Campaign xApp, traffic steering, and GS-lite stream processing engine\cite{osc_ric_apps}}}
&\makecell{\parbox{0.22\textwidth}{\tabitem onos-kpimon (KPM monitoring), onos-rsm (slice management), onos-mho (mobile handover for mobility management), onos-mlb (load balancing between cells), onos-pci (for managing PCI resources)}}
&\makecell{\parbox{0.22\textwidth}{\tabitem Besides the xApps provided by OSC, OAIC introduced their own KPI monitor and slice control xApps }}\\\hline

\makecell{\parbox{0.03\textwidth}{Lang. \\ Lic.}}
&\makecell{\parbox{0.22\textwidth}{\tabitem \textit{C/C++} \\\tabitem OAIPL1.1 }} 
&\makecell{\parbox{0.22\textwidth}{\tabitem \textit{Python}, \textit{Go} , and \textit{C/C++}\\\tabitem ALV2 mostly besides CCLA4I}}
&\makecell{\parbox{0.22\textwidth}{\tabitem \textit{Go} \\\tabitem ALV2}}
&\makecell{\parbox{0.22\textwidth}{\tabitem \textit{C/C++} \\\tabitem GAGPLV3}}\\

        \bottomrule
    \end{tabular}
\end{table}

\section{xApp Implementation-Driven Ambiguities Related to O-RAN Architecture} 
While standardization bodies define how the O-RAN architecture should be implemented to address various applications, there are some ambiguities observed while working on specific use cases. Here we focus on Beam Mobility Management (BMM-xApp) and Signaling Storm Detection xApps (SSD-xApp). The use cases related to those xApps are analyzed within O-RAN ALLIANCE's documents.

\subsection{Example RRM xApp - Beam Mobility Management}
\label{sec_BM}
One of the key technologies used in 5G NR is a Grid of Beams (GoB) beamforming. A UE is assigned to a specific beam (out of a static set) based on the downlink measurements of Reference Signal Received Power (RSRP). The measurements are typically carried using the Synchronization Signal Block (SSB), i.e., every 20 ms\cite{barati2020energy}. SSBs transmission for all beams lasts 5~ms. The main challenge, in this case, is when the UE moves fast. Under such conditions, the radio environment can rapidly change causing a radio link failure, e.g., due to signal blockage by the obstacles. To avoid such situations there is a need for AI/ML-assisted algorithms that utilize e.g., context information like UE location information, to infer future target beams possibly minimizing the number of beam reselections~\cite{ORANMassiveMIMOTR}.

\begin{figure}[htbp]
\centerline{\includegraphics[scale=0.5]{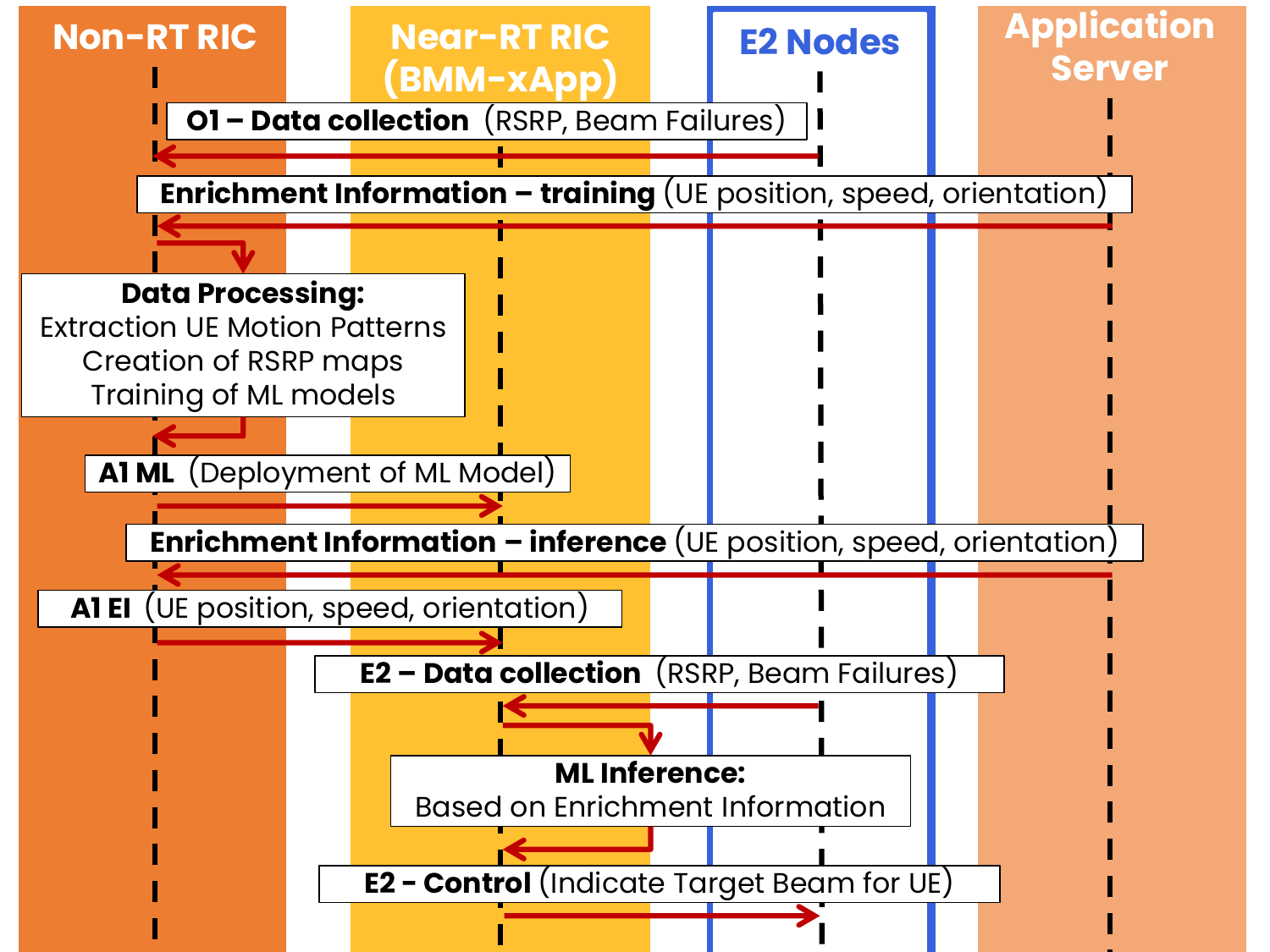}}
\caption{The information flow between the BMM-xApp, and other O-RAN entities}
\label{fig:bmm_oran}
\end{figure}

This is addressed by the use case "AI/ML-assisted Beam Selection Optimization" from O-RAN ALLIANCE \cite{ORANMassiveMIMOTR}. While its specification is somehow general up to now, i.e., O-RAN ALLIANCE specified utilized entities, e.g., Near-RT RIC, and interfaces, e.g., O1 and A1-ML, we had proposed the remaining elements, e.g., utilization of ML algorithm and data collected from E2 nodes.  
The information flow between the BMM-xApp, and other O-RAN entities is depicted in Fig.~\ref{fig:bmm_oran}. 
The main concept of this development is to perform the most extensive computations related to data analysis, and training of the ML model in the Non-RT RIC. The Near-RT RIC receives the pre-trained ML model and uses it for inference of target beams for the UEs. 
First, the O1 interface is configured to provide Non-RT RIC with users' RSRP measurements and beam failure statistics from E2 nodes. The beam failure statistics are used to monitor ML model accuracy, i.e., when the observed number of beam failures increases, the ML model re-training is triggered. 
RSRP measurements are used to create an RSRP map for each beam, following the Radio Environment Map (REM) concept~\cite{perez2015}. For this purpose EI, specifically: the position, speed, and orientation of each user, is obtained from the Application Server (specifically - the location server). 
The obtained data is being processed in the Non-RT RIC. First, the location information is analyzed to extract the UE Motion Patterns. They are, e.g., in the form of histograms that represent the probability of future UE speed and orientation while being in a particular location. A representative example can be a vehicular scenario. When users encounter a road intersection majority of them turn right, while only a few turn left. Next, the RSRP map is created, i.e., for each beam associated with a considered BS the spatial distribution of RSRP is created by aligning location information from the external Application Server, and RSRP collected from E2 Nodes. The alignment can be done through a comparison of the data if these are accurately timestamped. These RSRP maps capture specific radio environment characteristics, e.g., some beams can be blocked by obstacles in a particular location. Both UE Motion Patterns and RSRP maps represent the radio environment and are used to train ML models. The ML models can be trained according to different optimization goals, e.g., minimization of beam reselections while maintaining users' QoS, or SNR maximization. Reinforcement Learning (RL) can be used as it learns through interaction with the environment (wireless network)~\cite{Hoejoo2021}. After the training is finished the obtained ML model is transferred to Near-RT RIC via the A1 ML interface and deployed in the BMM-xApp to make inferences on target beams for UEs. To provide input to the deployed ML model, EI (specifically: location information) must be sent from the external Application Server to the BMM-xApp. This is done in a two-stage manner: first EI is sent to the Non-RT RIC, and next, it is forwarded to the Near-RT RIC through the A1-EI. In addition, the E2 interface is configured to collect information about the RSRP, and beam failures. First, the UE's localization is used in the ML inference performed by BMM-xApp, i.e., the target beam is selected. Secondly, the BMM-xApp monitors beam failures to validate the ML model performance. If too many beam failures occur it is a signal that the ML model is outdated. In such a case BMM-xApp can temporarily switch to the \emph{emergency} mode in which some analytical beam management procedure based on RSRP reports is performed (e.g.,~\cite{Abinader2021}) until a new ML model is provided from the Non-RT RIC.  

Recalling that this use-case is at its early stage of specification in O-RAN ALLIANCE, still, some implementation ambiguities are observed:
\begin{itemize}
    \item \textbf{Location information} is currently, not discussed within the O-RAN specifications, it is only mentioned as a specific type of EI message. However, it could be used by many xApps, and some of its aspects should be discussed within O-RAN ALLIANCE workgroups. The localization server should at last provide the following:
    \begin{itemize}
        \item \textit{Localization technique} that was used to obtain the location information, i.e., there are many localization techniques of significantly different accuracy, e.g., standard Global Navigational Satellite System (GNSS) receiver is characterized by an accuracy of 10 meters, while Real Time Kinematics (RTK) introduces only a centimeter-level error.
        \item \textit{Available measurements} that can be provided together with the user's position, e.g., user's speed, and bearing.
        \item \textit{Report intervals} should be possible to enforce or at least report. If the location information is provided only once per second the performance of BMM-xApp could be degraded as beam management can be triggered every 20~ms~\cite{barati2020energy}.
        \item \textit{Delay} can be introduced by passing the UE's localization information, required in Near-RT RIC, via Non-RT RIC as visible in Fig.~\ref{fig:bmm_oran}. A recently introduced Y1 interface between Near-RT RIC and Application Server can prevent such a potential bottleneck.
    \end{itemize}
    \item \textbf{Alignment of reported data in time}, i.e., precise timestamping of both RSRP and location information at the moment of measurement is crucial, e.g., for high-speed users which can travel a few meters during the time between the position was obtained and the EI was received in Non-RT RIC.
    \item \textbf{ML Modules}
    \begin{itemize}
        \item \textit{Deployment of ML Modules within O-RAN architecture} should be clarified. At the current stage of standardization, there are several options in SMO and Non-RT RIC architecture, where ML model training can be performed. As an example, training can be performed either by a vendor-dependent module, by dedicated rApp, within the rApp, or even outside of the Non-RT RIC and SMO. 
        \item \textit{A1 interface} specifications, at their current state, do not explicitly define ML Model service operations~\cite{ORANA1AP}.
    \end{itemize}
    \item \textbf{E2 interface} lacks actions related to beam management~\cite{ORANE2RC}, i.e., at this stage, it is unclear how BMM-xApp would enforce switching a particular user to the given beam.
\end{itemize}

\begin{figure}[htbp]
\centerline{\includegraphics[height=6.5cm, width=\textwidth]{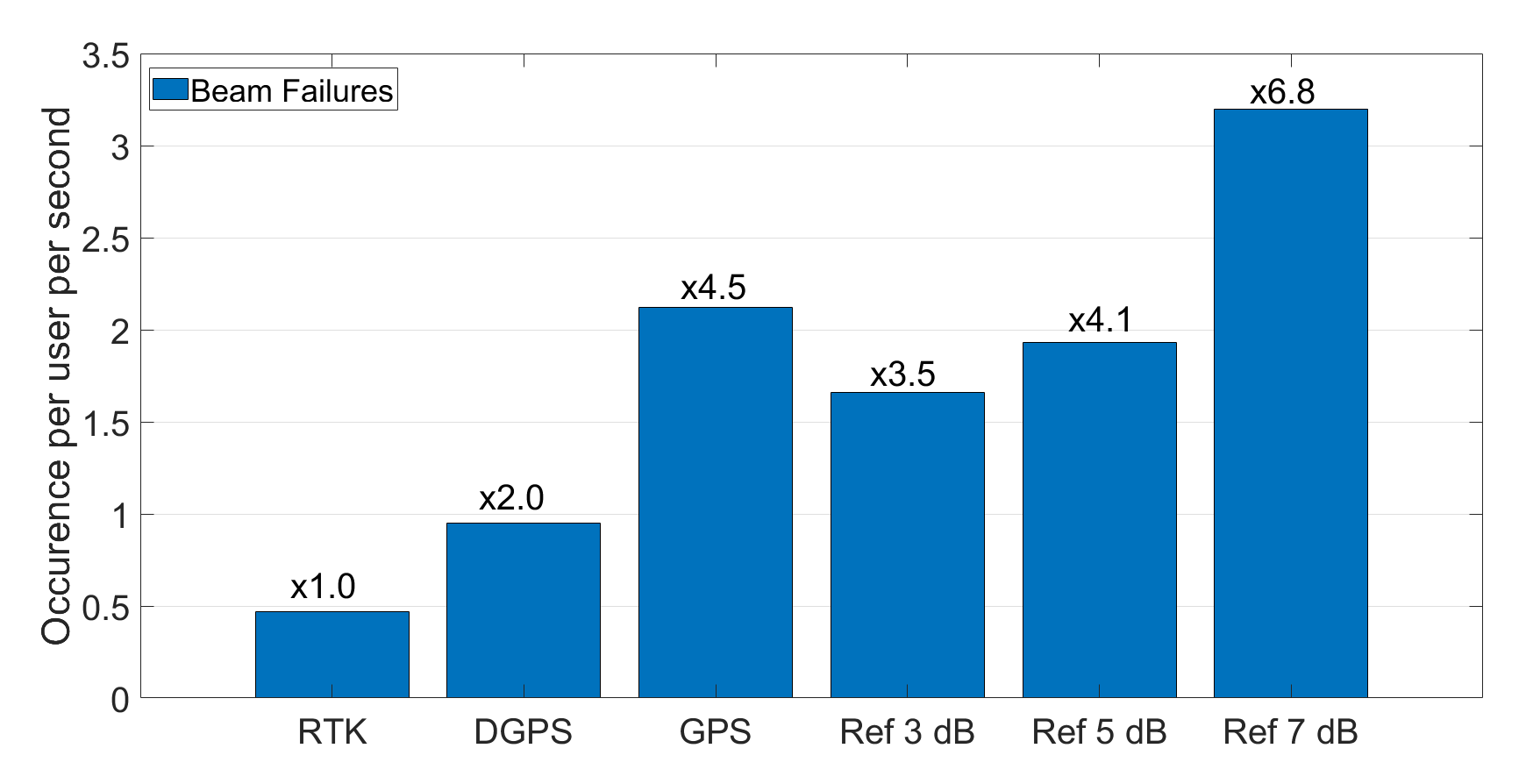}}
\caption{Number of beam failures per user, per second in the function of utilized localization technique.}
\label{fig:bf_loc}
\end{figure}

{\MH To highlight the importance of the quality of location information for the BMM-xAPP relying on the REM we have performed computer simulation studies in the scenario described in detail in~\cite{hoffmann2023beam}. The scenario considers a single Massive MIMO BS, operating at mmWaves frequency band, that supports 8 beams. Within this cell, we have placed 300 UEs moving upward with the speed of 25~m/s, to reflect the road scenario. We have tested the BMM-xApp following the optimization goal of minimization of beam reselections while avoiding beam failures under three localization techniques: RTK, Differential Global Navigation Satellite System (DGPS), and standard GPS. The standard deviations of their corresponding localization error are as follows~\cite{misra2006global}: 1~cm, 1~m,6~m, for RTK, DGPS, and GPS respectively. We have compared the results in terms of the observed number of beam failures per user, per second as depicted in Fig.~\ref{fig:bf_loc}. RTK provides almost perfect location information, but some beam failures occur due to channel variations. However, when additionally localization accuracy is degraded, more beam failures occur, i.e., compared to the RTK it is about 1.62 and 4.19 times more beam failures while utilizing DGPS, and GPS respectively. Thus, the information about the supported localization technique would be necessary for designing robust xApps.}

\subsection{Example Security xApp - Signalling Storm Detection}
The signaling storm attack is aimed at causing Denial of Service (DoS) in a network by occupying radio resources in a CP by an adversary or malfunctioning  device~\cite{Francois2015}. Such devices can persistently send control messages like registration requests, that will be rejected after validation in the CN, or can intentionally disconnect from the network after a successful registration. Such behavior is especially dangerous in the Internet of Things (IoT) networks. The IoT devices have low complexity, and as such can be relatively easily hacked by adversaries to flood networks with CP messages, e.g., adversaries can install on the IoT device software that will constantly restart the device triggering the registration procedure. It is important to notice that such a device will be authorized to connect to the network, and as such hard to be detected~\cite{Pavloski2019}. From this perspective, it is important to equip 5G networks with an intelligent mechanism that can detect the signaling storm as close to its origin as possible, possibly at the stage of RAN. After detection, further communications with malfunctioning devices should stop to prevent flooding the CN with CP messages. 

\begin{figure}[htbp]
\centerline{\includegraphics[height=8.0cm]{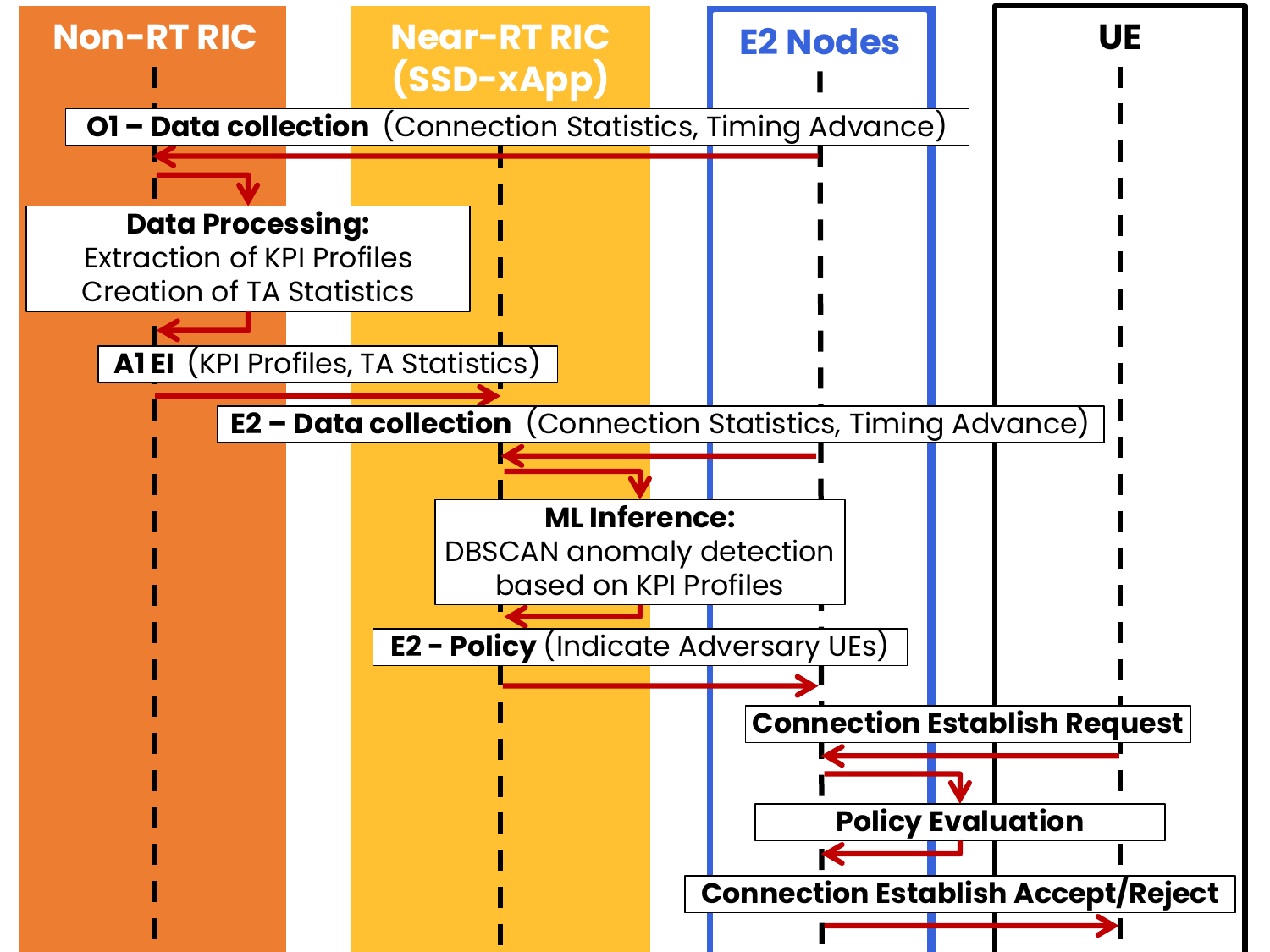}}
\caption{The information flow between the SSD-xApp, and other O-RAN entities}
\label{fig:ssd_oran}
\end{figure}

This xApp addresses a use case following requirements of O-RAN Signalling Storm Protection from~\cite{ORANUseCases} with a slight modification: here both attack detection and mitigation are integrated into a single SSD-xApp to reduce the amount of communication overhead. The O-RAN ALLIANCE specifies the high-level roles of the O-RAN entities and utilized interfaces for this use case. As the other details, e.g., ML method, and data exchanged with the E2 node, are missing here below we propose our solution keeping it fully compliant with O-RAN specification. 
The SSD-xApp utilizes the Timing Advance (TA) parameter being computed and exchanged at the early stage of the device's registration procedure (i.e., Msg2: Random Access Response~\cite{3gpp38321}). As this characterizes indirectly the distance electromagnetic wave travels between the UE and the BS it is difficult to be falsified. As such it can be utilized to filter malfunctioning devices creating an increased number of connection establishment requests, without interrupting CN functions, e.g., device authentication. The information flow between the SSD-xApp and other O-RAN entities is depicted in Fig.~\ref{fig:ssd_oran}. It starts with the configuration of the O1 interface to provide Non-RT RIC with connection statistics, including registration requests, RRC connection establishment requests, etc., and related TAs extracted from Msg2. This data is processed within the Non-RT RIC in order to produce the so-called Key Performance Indicator (KPI) Profiles~\cite{bodrog2016}. The KPI Profiles stores long-term statistics of a given KPI, e.g., the mean and standard deviation of a number of connection establishment requests over a period of the day. In addition, TA related to connection statistics is analyzed, e.g., in the form of histograms. The A1-EI is used to send KPI Profiles, and TA statistics observed over a long period in Non-RT RIC to the SSD-xApp residing in Near-RT RIC. This step should repeat periodically, e.g., twice a day, or on an event basis, e.g., when a high number of new UEs is deployed in a factory. 
The SSD-xApp obtains from E2 nodes temporal information about the connection statistics (e.g., number of connections establish requests over the last 5 minutes), and related TAs. Next, the SSD-xApp compares the long-term KPI Profile with temporal connection statistics computing the so-called anomaly values. It utilizes the unsupervised learning clustering algorithm Density-Based Spatial Clustering of Applications with Noise (DBSCAN) to detect the abnormal activity of users in the network, i.e., signaling storm. When the signaling storm is detected, the SSD-xApp analyses statistics of TA to produce a policy that will filter out connection establishment requests related to users associated with those TAs. The formulated policy is sent to the E2 Nodes via the E2 interface. Based on that policy the E2 Node can either accept or reject the connection establish requests sent by the UE by comparing their TA, with blacklisted TAs defined in the policy.


As with the BMM-xApp, also here some implementation ambiguities can be mentioned: 
\begin{itemize}
    \item \textbf{Resolution of TA} relies on the network configuration. A low resolution of TA will increase the number of devices having the same TA and potentially blocked. From this perspective it might be useful to provide the xApp with some extra historical information about the UE context from the CN registers, to distinguish an adversary from a legitimate user, e.g., historical channel state information, network identifiers, etc. 
    \item \textbf{Non-RT RIC} architecture is not clearly specified in terms of storage processing of EI~\cite{nonRTRIC}. In the case of KPI Profiles utilized by the SSD-xApp, it is unclear, whether there would be some dedicated vendor-dependent Non-RT RIC module for processing and storage of such xApp-provider-defined EI, or whether this functionality would be realized by some rApp.
    \item \textbf{E2 interface} policy service is not clearly defined within the O-RAN specifications~\cite{ORANE2RC}. It might happen that E2 Nodes would not support rejecting connection establish requests on the basis of the TA parameter.
\end{itemize}

{\MH To highlight the importance of the above-mentioned ambiguities, we have studied the potential impact of the TA resolution on the number of legitimate users that are being rejected from the network when adversary activity is detected. We are considering a simulation setup described in our previous work~\cite{hoffmann2023signaling}: a single cell of IIoT network of a 2~km radius, with 100 statically deployed legitimate IIoT sensors and 5 adversaries. Intervals between legitimate users' connection requests follow the exponential distribution with a rate parameter equal to 5 per hour. Each adversary performs on average 3 attacks per day consisting of 100 consecutive connection requests, send within the intervals of 5~s. Because the TA resolution depends on the utilized subcarrier spacing, here we have considered values proper for a 5G system: $15$, $30$, $60$, $120$, and $240$~kHz, respectively. The results are depicted in Fig.~\ref{fig:ta_scs}. It can be seen that for high values of subcarrier spacing, detection of adversary almost doesn't affect the performance of legitimate users, i.e., all their connection attempts are accepted. On the other hand, while utilizing low subcarrier spacing of $15$~kHz above 60\% of legitimate devices are rejected from the network because their TA is the same as the TA of the detected adversary. 
\begin{figure}[htbp]
\centerline{\includegraphics[height=6.5cm]{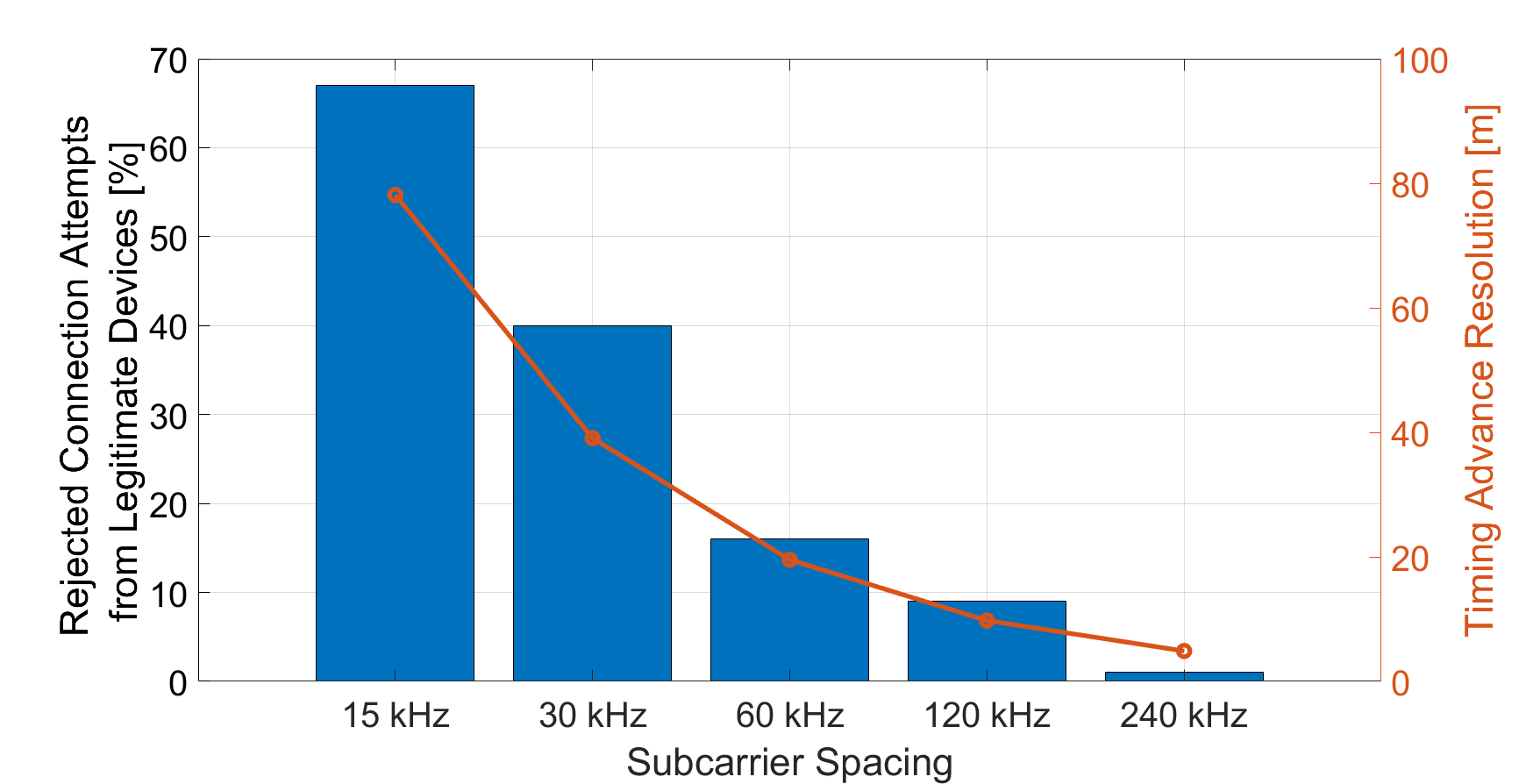}}
\caption{The ratio of rejected connection attempts from legitimate devices in the function of subcarrier spacing.}
\label{fig:ta_scs}
\end{figure}
} 


\section{xApp Implementation-Driven Conclusions Related to Implementation}
Contrary to the prior chapter, where we focused on the overall ambiguity related to xApp development, here we concentrate on various issues related to the detailed application implementation on selected open RIC platforms. We have selected for comparison two xApps - Traffic Steering xApp (TS-xApp) and QoS-Based Resource Allocator xApp (QRA-xApp), which consider use cases standardized by the O-RAN ALLIANCE specifications \cite{ORANUseCases}.  

{\AS The xApps have been deployed within the environment running on the virtual machine with the Ubuntu operating system (OS). It is based on the architecture packed in Kubernetes pods and Docker images. To ensure proper implementation of the xApps, the following virtual hardware requirements are obligated: a) processor with at least $2$ cores, b) Random Access Memory (RAM) with the size of min. $8$ GB, c) Read-Only Memory (ROM) with the size of min. $50$ GB, d) Ubuntu OS version 20.04.5 LTS.
}

\subsection{Traffic Steering xApp}
{\MH 
\begin{figure}[htbp]
\centerline{\includegraphics[height=7.0cm]{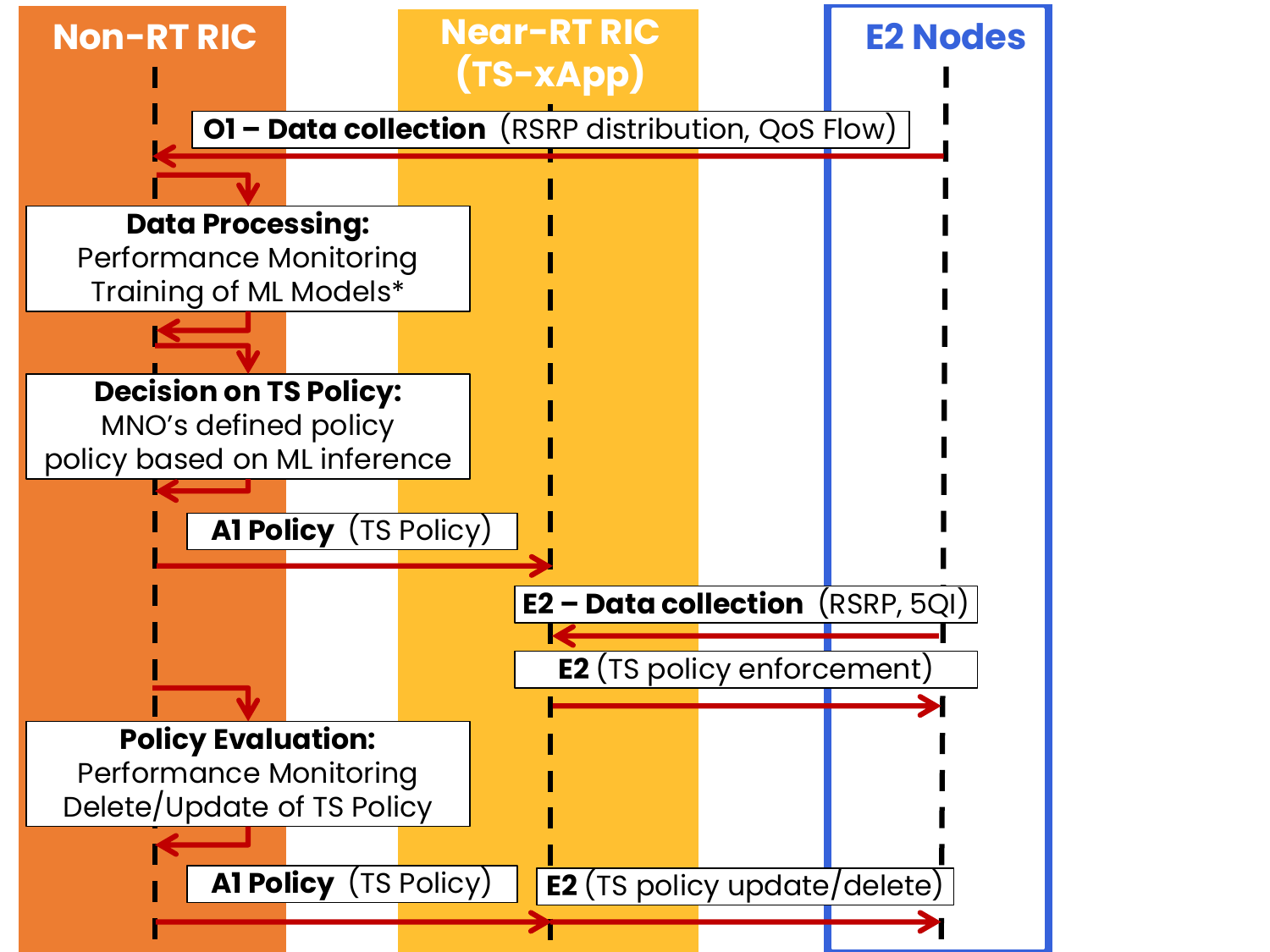}}
\caption{The information flow between the TS-xApp, and other O-RAN entities}
\label{fig:ts_oran}
\end{figure}
}

TS-xApp addresses the use case \#5: O-RAN Traffic Steering from \cite{ORANUseCases}. It provides the ability to dynamically switch mobile users between cells available in the access network. The purpose of such a mechanism is to manage the current mobile traffic in order to ensure the high performance of the radio system. Depending on actual needs, the MNO can realize various TS targets such as guaranteeing equal traffic load for all nodes (load balancing), separating users with different Quality-of-Service (QoS) demands (service-based association), supporting the reduction of energy consumption, and many others.

In the TS xApp, the user association is performed through the E2 Interface by using an O-RAN-defined handover control mechanism. The decisions about switching users among cells are done based on the RSRP distribution reports received through the E2 Interface, and policies that are sent by the Non-RT RIC through the A1 interface. Inside these policies, the rules, which indicate preferred and forbidden cells for a particular UE, can be found. The preferences can be oriented to users assigned to a particular slice (slice-oriented approach) or having strictly specified identification (user-centric approach). The A1 policies are exchanged between Non-RT RIC and TS xApp in the form of JavaScript Object Notation (JSON) files, which are prepared according to the schema of the “Traffic Steering Preferences” type standardized by the O-RAN ALLIANCE \cite{ORANA1TD, ORANUseCases}. The information flow is depicted in Fig. \ref{fig:ts_oran}.

{\AS TS-xApp has been integrated with the SD-RAN environment provided by the ONF; it is able to handle connections with the $\mu$ONOS RIC components of the SD-RAN. Furthermore, it can interpret received E2 and A1 messages properly and suggest (to RIC) performing adequate handover operations, the results of which are reflected in the RAN Simulator. The source code of the xApp can be found in \cite{rimedo_ts}. In Tab.~\ref{tab_results_ts}, the results for the TS xApp have been attached. To observe the performance of the system, the considered, intentionally-simple scenario consisted of two one-cell base stations and a single UE terminal, which was moving between the locations of both BSs. Within the tests, three different UE-oriented policies were enforced. Those policies indicated the preferences for connection handling with the user by a particular cell - PREFER, AVOID, and FORBID. Cells marked in a policy with these labels were recognized by the UE as cells, by which the UE should, should not, and must not be served, respectively. Thus, referring to Tab.~\ref{tab_results_ts}, it can be observed that when the connection between the user and cell is marked with the PREFER label, this link is handled for $75\%$ of the observation time. The AVOID mark causes the opposite result -- the UE is served by such a cell for $25\%$ of the observation time. Next, the FORBID label resulted in not serving the user by a given cell. Finally, in the case where there was no policy enforced for the TS xApp, the UE was associated with a cell based on the RSRP report. Thus, it was noticed the user was served for $50\%$ of the observation time by one cell and $50\%$ by another.

\begin{table}[t]
\label{tab_results_ts}
\centering
\caption{Association of the UE within the network by the TS xApp according to different policies}
\begin{tabular}{|c|c|c|c|c|}
\hline
\multirow{3}{*}{POLICY NAME} & \multicolumn{4}{c|}{USER ASSOCIATION TIME PART {[}\%{]}}                                                                                          \\ \cline{2-5} 
                             & \multicolumn{2}{c|}{ENFORCED FOR $1^\text{st}$ CELL}                               & \multicolumn{2}{c|}{ENFORCED FOR $2^\text{nd}$ CELL}          \\ \cline{2-5} 
                             & \multicolumn{1}{c|}{$1^\text{st}$ CELL} & \multicolumn{1}{c|}{$2^\text{nd}$ CELL} & \multicolumn{1}{c|}{$1^\text{st}$ CELL} & $2^\text{nd}$ CELL \\ \hline
NONE                         & \multicolumn{1}{c|}{$50$}                & \multicolumn{1}{c|}{$50$}                & \multicolumn{1}{c|}{$50$}                & $50$                \\ \hline
PREFER                       & \multicolumn{1}{c|}{$75$}                & \multicolumn{1}{c|}{$25$}                & \multicolumn{1}{c|}{$25$}                & $75$                \\ \hline
AVOID                        & \multicolumn{1}{c|}{$25$}                & \multicolumn{1}{c|}{$75$}                & \multicolumn{1}{c|}{$75$}                & $25$                \\ \hline
FORBID                       & \multicolumn{1}{c|}{$0$}                 & \multicolumn{1}{c|}{$100$}               & \multicolumn{1}{c|}{$100$}               & $0$                 \\ \hline
\end{tabular}
\end{table}
}
\subsection{QoS-Based Resource Allocation xApp}

{\MH 
\begin{figure}[htbp]
\centerline{\includegraphics[height=8.0cm]{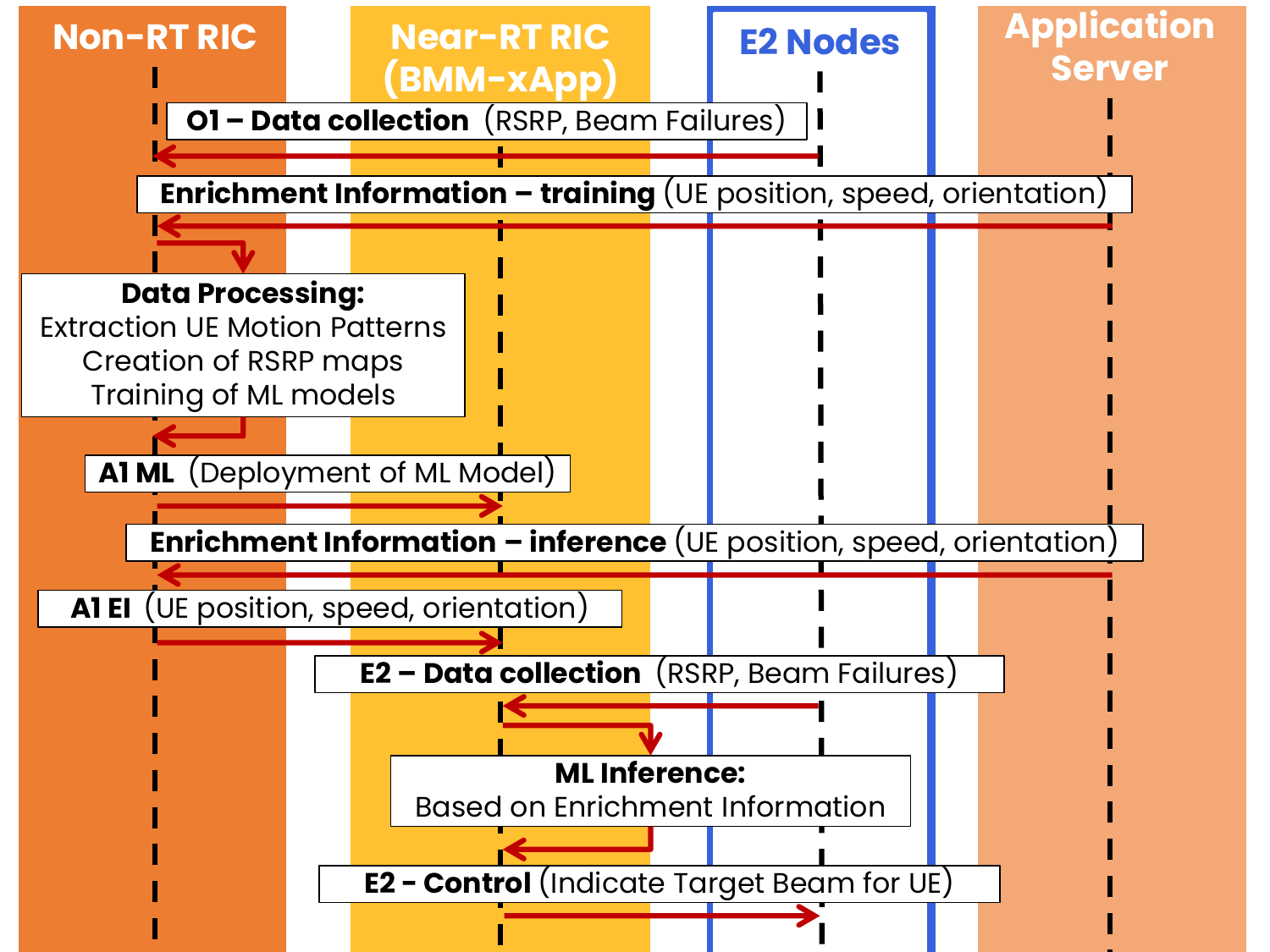}}
\caption{The information flow between the QRA-xApp, and other O-RAN entities}
\label{fig:qra_oran}
\end{figure}
}

QRA-xApp addresses use case no. 8: O-RAN QoS Based Resource Optimization from \cite{ORANUseCases}. It is responsible for splitting radio resources in the form of Physical Resource Blocks (PRBs) among available slices within the network. With the QRA xApp, the MNOs are able to manipulate the radio resources allocated by the scheduler to manage the networks' performance by allocating more PRBs for high-performance slices (e.g., Mobile Broadband - MBB) and simultaneously reducing the number of resources for slices demanding low data rate (e.g. Voice services).

This allocation of radio resources is done in order to meet the SLA targets defined inside policies (passed to xApp in the form of JSON files by the Non-RT RIC through the A1 Interface), by basing on measurement reports received through the E2 Interface for a particular slice served by some gNB. The SLA targets are specified in the A1 policy file as a throughput rate expressed in $\left[\text{bps}\right]$, which can be translated to the number of needed PRBs (and vice versa) by taking into account current propagation conditions for a slice (e.g. SNR/RSRP distribution, number of active UEs, service/slice types, etc.). This group of SLA targets specified inside A1 policies consists of UE- or slice-oriented parameters such as Guaranteed and Maximum Throughput per Slice, Maximum Throughput per UE, Maximum Number of UEs per Slice, etc. The used shape of A1 policies has been defined by the O-RAN ALLIANCE as the schema of policy type called “SLA Target” \cite{ORANA1TD, ORANUseCases}. The information flow between the involved entities is depicted in Fig. \ref{fig:qra_oran}.

{\AS QRA-xApp, similar to TS-xApp, has been integrated, and tested using the ONF's SD-RAN environment. Its working has been verified for the scenario using two one-cell mobile access nodes again, but at least six user terminals, which constantly move together from one base station to another. The QRA xApp provides connections with the SD-RAN's $\mu$ONOS RIC components. Thanks to the correct interpretation of received E2 and A1 messages, the xApp performs adequate resource-allocation-related operations, the results of which (delivered via RIC to E2 nodes) could be visible in real-time mode in the form of terminal logs. 

\begin{table}[b]
\centering
\caption{Radio resource allocation for different schemas within the QRA xApp}
\label{tab_results_qra}
\begin{tabular}{|c|c|ccc|}
\hline
\multirow{2}{*}{UE ID} & \multirow{2}{*}{$5$QI} & \multicolumn{3}{c|}{BANDWIDTH PART {[\%]}}                         \\ \cline{3-5} 
                       &                      & \multicolumn{1}{c|}{EQUAL} & \multicolumn{1}{c|}{PREFER-$3$} & RESERVE \\ \hline
$1$                      & $1$                    & \multicolumn{1}{c|}{$12.5$}  & \multicolumn{1}{c|}{$6.25$}     & $5$       \\ \hline
$2$                      & $2$                    & \multicolumn{1}{c|}{$12.5$}  & \multicolumn{1}{c|}{$6.25$}     & $10$      \\ \hline
$3$                      & $4$                    & \multicolumn{1}{c|}{$25$}    & \multicolumn{1}{c|}{$12.5$}     & $40$      \\ \hline
$4$                      & $2$                    & \multicolumn{1}{c|}{$12.5$}  & \multicolumn{1}{c|}{$6.25$}     & $10$      \\ \hline
$5$                      & $3$                    & \multicolumn{1}{c|}{$25$}    & \multicolumn{1}{c|}{$62.5$}     & $30$      \\ \hline
$6$                      & $1$                    & \multicolumn{1}{c|}{$12.5$}  & \multicolumn{1}{c|}{$6.25$}     & $5$       \\ \hline
\end{tabular}
\end{table}

In Tab.~\ref{tab_results_qra}, the results for the QRA xApp have been attached. The considered scenario consisted of two one-cell base stations and six UE terminals, which were moving simultaneously between the locations of both BSs. Each user served within the network could belong to a different slice. In a particular slice, all UEs connected to a specific network cell and using the same service type (denoted by the $5$G QoS Identifier -- $5$QI) were grouped. Within the tests, four service types ($5$QI equal to $1$, $2$, $3$, or $4$) and three different schemas of radio resource allocation (EQUAL, PREFER-X, and RESERVE) have been taken into account. According to the EQUAL approach, all available PRBs were divided among existing slices equally. Next, the PREFER-X schema (where X is the number indicating the service type, i.e., the $5$QI, of a particular slice -- $1$, $2$, $3$, or $4$) shares all the resources among the slices in the ratio of 5:1 for ones with "preferred" service type ($5$QI) to the rest of them. Finally, the RESERVE approach divides all PRBs within the cell among available slices in the ratio of $5$X, where X is the number that indicates the service type of a given slice ($5$QI). Thus, for our scenario with four different service types ($1$, $2$, $3$, and $4$), the ratio of sharing the resources for RESERVE schema is equal to $5$:$10$:$15$:$20$. 
}

\subsection{Implementation of xApps - Challenges and Limitations}
In Table~\ref{tab:challenges} we summarize the challenges faced during the development, deployment, and testing of xApps using different platforms.
\begin{table}[h]
\centering
\caption{Challenges faced during xApp development and testing.}
\label{tab:challenges}
\begin{tabular}{p{.1\linewidth}M{.9\linewidth}}
        \toprule
\textbf{Aspect} & \textbf{Challenges}\\\midrule
Simulator
&\tabitem Available RAN simulators do not provide complete functionality needed to test different specific practical scenarios (e.g., different network size, base station capabilities, network operation duration, etc.)\\\midrule

Conflicts
&\makecell{\parbox{0.9\textwidth}{\tabitem Absence of conflict mitigation units prevents testing the operation of multiple xApps working simultaneously\\
\tabitem Multiple A1 policies that could be turned on simultaneously should be verified against each other beforehand}}\\\midrule

SDK/API & \makecell{\parbox{0.9\textwidth}{\tabitem Standard compliance: base stations or simulators do not provide the functions or parameters needed for complete O-RAN functionality implementation\\
\tabitem Abstraction of O-RAN messages that implement certain functionalities (e.g., RSRP monitoring, handover control, etc.) would simplify xApp development process\\
\tabitem Interoperability between components like simulators and RICs\\ 
\tabitem Exemplar xApps should be provided and they should cover the functionality of the platform as much as possible}}
\\
        \bottomrule
    \end{tabular}
\end{table}

\section{Challenges for O-RAN/Incentives to O-RAN Triggered Research}
Following the discussion on xApp/rApp implementation and deployment issues, in this section, we try to identify the key challenges that appear on the Open RAN development path.
\subsection{Challenge A: The Need for Intelligent Conflict Management}
Intelligent RAN control functions enabled in the Near-RT RIC with the introduction of xApps allow for flexibility in the adaptation of network operation characteristics. While implementing a single application, there is no need for any mechanism responsible for conflict management; what is necessary is only the subscription functionality, so that the particular xApp or rApp can request access to specific parameters or metrics through standardized service models. On the other hand, having multiple xApps/rApps, developed by various third-party providers, working simultaneously in RICs will inevitably lead to conflicts between control actions affecting the E2 Nodes finally. Thus, the incorporation of two (or more) xApps/rApps immediately entails the need for stable and precise solutions for conflict management \cite{Adamczyk2023}.
The xApp/rApp developer has to be aware of the applied policy in case of any prospective conflicts - if any priority or hierarchy between the application will be applied and how it may or will impact the functioning of the application. Based on our implementation experience, it is one of the key challenges that have to be effectively solved to enable reliable xApp provisioning. 

\subsection{Challenge B: Security}
Another critical point that was immediately observable during the implementation of the xApps/rApps is related broadly to Open rAN security - both on the architectural side and from the perspective of xApp/rApp delivery by the third party. 
When talking about the security of an O-RAN architecture, one should note that the attack surface is expanded as compared to the standard radio segment of a mobile communication network. This surface contains ``traditional'' attacks related to the omnipresent radio transmission medium, cyberattacks related to virtualization (softwarization) of RAN functions, i.e., attacks on xApps, rApps, and edge Artificial Intelligence (AI) algorithms residing in O-RAN and Multi-access Edge Computing entity (MEC), as well as attacks related to O-RAN interfaces.

The O-RAN specification and \emph{openness} of the radio interface poses challenges for the entire network security. Inadequately defined and poorly secured O-RAN applications and interfaces (including the front-haul interface, O1, O2, A1, and E2) can potentially be targets of attacks. Attackers can utilize these new open interfaces to attack the system, which could lead to a denial of service, data tampering, or data leaking, all of which have an indirect impact on the system's security. Each O-RAN interface and function may be subject to different threats, and each threat will have a particular impact, thus, for each threat, specific security measures and solutions must be used for all aspects and assets \cite{Shen_2022}. Finally, AI and ML algorithms residing at the network edge  (a consequence of the ML-as-a-Service paradigm for 5G/6G networks) become a target of a new type of attack - attacks on AL/ML. These threats can be classified as (i) \emph{poisoning attacks} manipulating the data or the learning algorithm in the model training phase, (ii) \emph{evasion attacks} aiming at the inference stage (test phase) based on the previously learned model, and (iii) \emph{inference attacks} aiming at recovering the training data and/or their labels, discovering the model architecture and its parameters \cite{Benzaid2020}.

At the same time, O-RAN architecture can be used to increase security in radio access networks because it allows for running xApps in Near-RT RIC, which can be developed to continuously monitor and analyze security threats and protect RAN from malicious and illegal access to network segments. It makes it possible to detect threats much faster before they affect the operation of the entire network. xApps can be developed for specific types of threats in a given network that can be detected closer to their occurrence. AI/ML algorithms can also be developed to improve security, e.g., by detecting various kinds of anomalies in radio traffic. Future research should aim at developing such xApps for O-RAN security despite expanded attacks surface.  

\subsection{Challenge C: The Need for Complete Automation {\AK and Testing} Procedures}
Another challenge that was raised immediately during the implementation of all the applications discussed above, is the stringent need for broad automation of the whole process of xApp/rApp delivery, testing, and deployment. As at the current stage the applications can be tested, verified, and installed manually, it is impossible to keep this stage in the future. Thus, based on the gained experience we claim that one of the key challenges at the current stage of O-RAN development is the lack of automation process related to testing and installation of the xApps/rApps on the RIC platforms. {\AK The template-based approach for xApp and rApp development is discussed in \cite{Kliks2023}. A general automated, distributed and AI-enabled testing framework has been presented in \cite{Tang2023}, with the purpose to test AI models deployed in O-RAN.}

This currently requires manual integration of the application every time a new one is to be deployed. There is no unified way how to smoothly introduce new/upgraded xApps to the system which consumes the resources of both, the providers and the operators/customers. The xApp providers utilize the resources for this purpose instead of focusing on developing and improving the algorithms, while the customer/receiver unnecessarily utilizes time to have to manually check that the xApp performs according to its design.

\subsection{Challenge D: Portability}
Yet another topic that yields currently cumbersome tasks is the portability of xApps/rApps between RIC platforms. What has been heavily experienced is that having the same core algorithm requires significant manual integration work to deploy it as an xApp on one RIC, with a more or less similar amount of work, when putting the same algorithm onto xApp for a different RIC. There are several reasons influencing this situation. First, there are different maturity levels of the various commercial and open-source RIC platforms, where each focuses on a different aspect. Second, the standardization of the RICs, as well as E2 and A1 interfaces is not yet mature enough to have a clear implementation guide for the vendors. And finally, there is a lack of a standard for SDK/API/CDK such that the xApp/rApp could be ported from one RIC to another with minimal intervention to the packaging of the xApp. 

Due to the above, when having an algorithm, the xApp developer needs first to get up to speed with the RIC platform itself and accompanied SDK, to surround the xApp with the proper interfacing. There is yet another aspect to it, which is not directly related to the RIC platform itself, but rather to the corresponding E2 nodes, which it works with. It relates to the integration of the RIC with the particular RAN software which may utilize a different set of, e.g., E2 service models or different versions of the same E2 service model. In such a setup, the xApp may not get all the required parameters from the E2 node which the RIC platform works with. This requires modification in the xApp itself so that the algorithm takes into account either fewer parameters or different parameters compared to a different RIC-CU-DU constellation.

\subsection{Challenge E: Ambiguity in Implementation - Processing Resources Optimization}
Finally, from the perspective of xApp/rApp functionality design and testing, the final challenge is related to the ambiguity in implementation. While the O-RAN ALLIANCE defines use cases with examples of messages exchanged between nodes, the xApp/rApp developers should have freedom of implementation limited only by the interface specification. Only in this case long-term development and improvement of applications are possible. It will resemble a market where various products (applications) can compete and the best or the most suitable (for a given network) solution can be implemented. As an example: the BMM-xApp, as described in Sec. \ref{sec_BM}, can be implemented using both the ML modules in Non-RT RIC and xApp in Near-RT RIC. However, similar results, i.e., a decision of a beam reselection sent to gNB, can be obtained by a single rApp, xApp, or a combination of rApp and xApp. It is possible that the various solutions will use different sets of measurements for learning purposes.
The problem becomes even more significant while considering a use case not considered by the O-RAN specification. In order to be able to implement such an xApp/rApp sufficient freedom for developers is required. This shows that the set of parameters exposed on interfaces should be as broad as possible. On the other hand, each Application should be constantly monitored for the amount and type of information exchanged on the interfaces. Additionally, a responsible RIC (directly or indirectly first to get support from SMO) should take care of the computational and storage resources required by a given application. If unlimited freedom is given to developers, it is possible that the application will poorly scale with, e.g., the number of UEs or operation time. 
If the limit is reached, the application should be \emph{killed} and somehow reported to the community and developers.

\section{Conclusion}
Open RAN as the technology is still in one of its initial phases of development. Much effort is put toward a precise and adequate definition of various standards, reflecting different aspects of the Open RAN community. Moreover, from a scientific perspective, numerous projects and activities have recently started that target many vivid and important problems related to the fair functioning of the complete open system. However, the whole process should also take into account the experience gained during initial implementation experiments and deployments. In this paper, we have described the lessons learned during the practical implementation of some xApps, selected based on the indications originating from the O-RAN ALLIANCE documents.  It has been shown that from the perspective of xApp/rApp algorithmic design, there is still a bit of ambiguity in the overall architecture. It limits the scope of perspective investigation of the proposed solutions. Next, in-detail implementation of the selected applications led to the identification of the key modifications and adjustments that could potentially improve the impact of the open-source RIC platforms. Finally, the overall discussion on the xApp development and deployment process allowed us to identify precisely five key challenges that must be handled in the near future. As these challenges impact various aspects of the open RAN concept, it is evident that joint efforts from academia, standardization body, and industry are necessary. We claim that without tight cooperation between these {\AK three sectors}, the further development of the open, disaggregated, flexible, and modular radio access networks will be limited.

\section*{Acknowledgments}
This work was funded in part by the National Centre for Research and Development in Poland within the 5GStar project CYBERSECIDENT/487845/IV/NCBR/2021 on ``Advanced methods and techniques for identification and counteracting cyberattacks on 5G access network and applications''. 




%

\bibliography{bibliografia}
\bibliographystyle{IEEEtran}

\vfill

\end{document}